# Measure-resend authenticated semi-quantum key distribution with single photons

Chun-Hao Chang[1], Yu-Chin Lu[2], and Tzonelih Hwang[*]

[1, 2, *] *Department of Computer Science and Information Engineering, National Cheng Kung University, No. 1, University Rd., Tainan City, 70101, Taiwan, R.O.C.*

[1] j7itgo@gmail.com

[2] m82617m@gmail.com

[*] hwangtl@csie.ncku.edu.tw (corresponding author)

## Abstract

Yu et al. and Li et al. have proposed the measure-resend protocols of authenticated semi-quantum key distribution (ASQKD). A new measure-resend ASQKD protocol is proposed in this paper, which requires a lower burden of quantum resource, needs fewer bits of the pre-shared key, and even provides better qubit efficiency than their protocols. The security proof shows the robustness of the proposed protocol under the collective attack.

## 1 Introduction

The quantum key distribution (QKD) is one important research topic of quantum cryptography, enabling a secret key to be distributed on the quantum channel among participants. In 2007, Boyer et al. [1] first proposed the protocol of semi-quantum key distribution (SQKD), in which Bob is a classical participant who can only perform three out of the following four operations: (1) measure qubits in The z-basis, (2) generate qubits in The z-basis and send, (3) reorder qubits, and (4) reflect qubits. In particular, the measure-resend SQKD in [2] allows the classical user to perform (1), (2), and (4) operations, together with the use of a classical authentication channel. In 2014, Yu et al. [3] proposed two authenticated semi-quantum key distribution (ASQKD) protocols, in which maximally entangled Bell states and pre-shared keys are used. They do not need the classical authentication channels, but a public classical channel to announce information during key distribution is still required. In 2016, Li et al. [4] proposed two ASQKD protocols, in which though Bell states are still used, the classical channel is removed[1].

The proposed protocol has the following features: First, low burden of the quantum resource, i.e., only single photons are used instead of maximally entangled Bell states. Second, the required pre-shared keys are reduced: two pre-shared keys are used. One is for the position of decoy photons, and the other is for choosing a hash function from the universal hash functions [5]. The total number of pre-shared keys are $n + 2m$ bits. Third, the security proof is given, which shows the proposed protocol has robustness under the collective attack. Finally, when there is no eavesdropping being detected, all the pre-shared keys can be recycled securely.

The rest of the paper is organized as follows: Section 2 illustrates the proposed ASQKD protocol. Section 3 presents the security analyses. Section 4 shows the comparison with Yu et al.'s and Li et al.'s measure-resend protocols. Section 5 concludes this paper.

## 2 Proposed measure-resend ASQKD protocol

The proposed protocol (shown in Fig. 1) uses two-step quantum communication. The quantum channel is assumed to be free from error and noise. Alice has full quantum capabilities while Bob has only limited quantum capabilities. Assume Alice would like to distribute a key to Bob. In the beginning, Alice and Bob both have pre-shared secret keys $K_1$, $K_2$ and a pool of universal hash functions. The length of $K_1$ is $n + m$ bits while the length of $K_2$ is $m$ bits.

---

[1] A 1-bit authentication message Alice used to inform Bob is still needed.

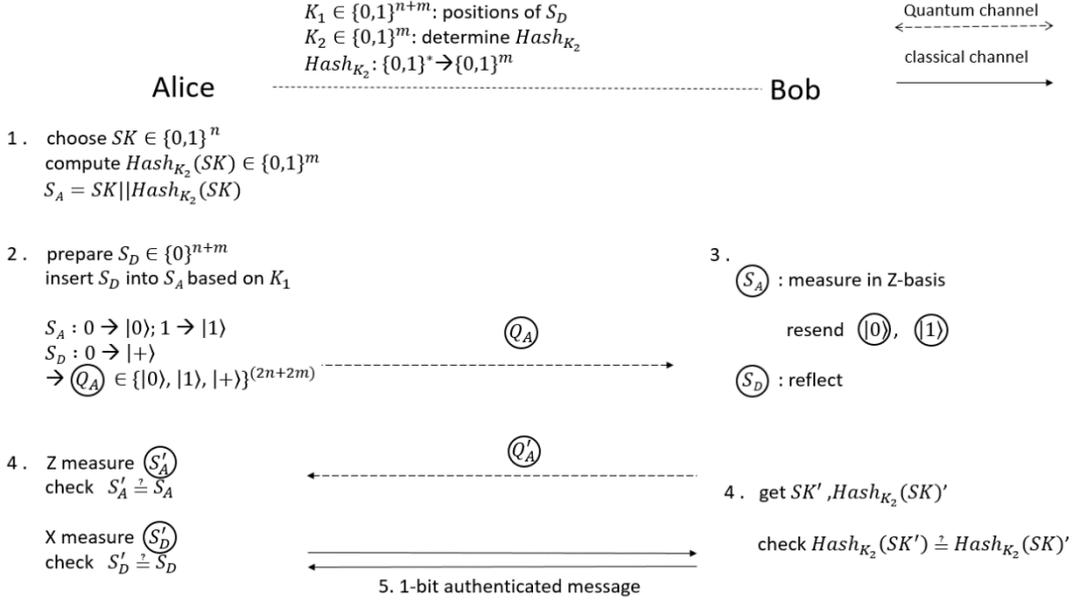

**Fig. 1.** The proposed ASQKD protocol

**Step 1.** Alice determines a key, $SK$, which is a binary string of length $n$, to be distributed to Bob. She chooses a hash function from the pool of universal hash functions based on the pre-shared key $K_2$. Then, $SK$ is hashed by the selected hash function to get an $m$-bit hash value. The binary string, $SK$, and the corresponding hash value are concatenated to become $S_A$, whose length is $n + m$ bits.

**Step 2.** Alice generates a string $S_D \in \{0\}^{n+m}$. $S_D$ is inserted into $S_A$ based on the pre-shared key $K_1$ (If the $i$-th bit of $K_1 = 0$, the $i$-th bit of $S_D$ is inserted in front of the $i$-th bit of $S_A$. Otherwise, it is inserted behind). Thus a binary string of length $2n + 2m$ bits is produced. Then, according to this binary string, Alice generates a sequence of single photons named $Q_A$ with the following rules: For $S_A$, 0 is encoded to $|0\rangle$, and 1 is encoded to $|1\rangle$. For $S_D$, all bits of $S_D$ are encoded to $|+\rangle = \frac{1}{\sqrt{2}}(|0\rangle + |1\rangle)$. Finally, she sends $Q_A$, one bit at a time, to Bob via a public quantum channel.

**Step 3.** According to the pre-shared key $K_1$, Bob can distinguish bits in $S_A$ from bits in $S_D$. For each received photon belonging to $S_A$, Bob measures it in The z-basis and resends the single photon in The z-basis based on the result of measurement (If the result of the measurement is 0 then resend $|0\rangle$, else resend $|1\rangle$). If the received photon belongs to $S_D$, and Bob reflects it without any disturbance. Since Bob has no delay line to store photons, each time he has to process one photon and send it back to Alice.

**Step 4.**

**Alice:** Upon receiving photons from Bob, Alice measures these photons. For those photons belonging to $S_A'$, Alice measures them in The z-basis. For photons belonging to $S_D'$, Alice measures them in The x-basis. If the binary strings of $S_A'$ and $S_D'$ from the measurement result are not the same as the original $S_A$ and $S_D$, there must be an eavesdropper's attack during the process.

**Bob:** In the meantime, while Bob gets $SK'$ and its corresponding hash value $Hash_{K_2}(SK)'$ from the measurement result in Step 3, he puts $SK'$ into the hash function and compares $Hash_{K_2}(SK')$ with $Hash_{K_2}(SK)'$. If $Hash_{K_2}(SK')$ is equal to $Hash_{K_2}(SK)'$, Bob can assure that he gets the distributed key sent by Alice. Otherwise, there must be an eavesdropper's attack.

**Step 5.** After the checks in Step 4, Alice and Bob will both send a 1-bit authenticated message to inform the other. If all checks are passed, the used pre-shared keys will be recycled. Otherwise, the result will be abandoned, and the used pre-shared keys are discarded.

## 3 Security analyses

In this section, we will analyze the security of the proposed protocol and prove that the proposed protocol is robust [1], which means when Eve attacks the protocol and gets some useful information, she will be detected with a non-zero probability. In this analysis, we assume Eve can fully control the quantum channel and has unlimited computational power, and she can perform the collective attack [6]. To perform the collective attack, Eve first prepares a quantum state $|E_i\rangle$, and performs a unitary operator $U_i$ on the joint state $|q\rangle \otimes |E_i\rangle$, where $|q\rangle$ is the qubit transmitted between participants, and $|E_i\rangle$ is Eve's prepared quantum state. Eve can use different quantum state $|E_i\rangle$ and unitary operators $U_i$ on the qubits which are sent from Alice or sent from Bob. But she needs to use same $|E_i\rangle$ and $U_i$ in every round of attack.

**Theorem 1.** The proposed protocol is robust under the collective attack, i.e., if there is no error, the protocol does not leak any information about the pre-shared key and the newly shared key.

*Proof.* We denote the unitary operator $U_1$ which Eve used to attack the qubits sent from Alice to Bob by:

$$U_1|00\rangle|E_1\rangle = a_{A1}|00\rangle|f_{A1}\rangle + a_{A2}|01\rangle|f_{A2}\rangle + a_{A3}|10\rangle|f_{A3}\rangle + a_{A4}|11\rangle|f_{A4}\rangle$$

$$U_1|01\rangle|E_1\rangle = b_{A1}|00\rangle|g_{A1}\rangle + b_{A2}|01\rangle|g_{A2}\rangle + b_{A3}|10\rangle|g_{A3}\rangle + b_{A4}|11\rangle|g_{A4}\rangle$$

$$U_1|10\rangle|E_1\rangle = c_{A1}|00\rangle|h_{A1}\rangle + c_{A2}|01\rangle|h_{A2}\rangle + c_{A3}|10\rangle|h_{A3}\rangle + c_{A4}|11\rangle|h_{A4}\rangle$$

$$U_1|11\rangle|E_1\rangle = d_{A1}|00\rangle|k_{A1}\rangle + d_{A2}|01\rangle|k_{A2}\rangle + d_{A3}|10\rangle|k_{A3}\rangle + d_{A4}|11\rangle|k_{A4}\rangle,$$

where $|E_1\rangle$ is Eve's prepared quantum state and $|f_{Ai}\rangle, |g_{Ai}\rangle, |h_{Ai}\rangle, |k_{Ai}\rangle$ are Eve's quantum states after the attack.

Because Alice will only send four different qubit pairs: $|0+\rangle, |+0\rangle, |1+\rangle,$ or $|+1\rangle$ in the protocol, we can denote these situations by:

$$U_1|0+\rangle|E_1\rangle = \frac{1}{\sqrt{2}}(a_{A1}|00\rangle|f_{A1}\rangle + a_{A2}|01\rangle|f_{A2}\rangle + a_{A3}|10\rangle|f_{A3}\rangle + a_{A4}|11\rangle|f_{A4}\rangle)$$
$$+ \frac{1}{\sqrt{2}}(b_{A1}|00\rangle|g_{A1}\rangle + b_{A2}|01\rangle|g_{A2}\rangle + b_{A3}|10\rangle|g_{A3}\rangle + b_{A4}|11\rangle|g_{A4}\rangle)$$

$$U_1|+0\rangle|E_1\rangle = \frac{1}{\sqrt{2}}(a_{A1}|00\rangle|f_{A1}\rangle + a_{A2}|01\rangle|f_{A2}\rangle + a_{A3}|10\rangle|f_{A3}\rangle + a_{A4}|11\rangle|f_{A4}\rangle)$$
$$+ \frac{1}{\sqrt{2}}(c_{A1}|00\rangle|h_{A1}\rangle + c_{A2}|01\rangle|h_{A2}\rangle + c_{A3}|10\rangle|h_{A3}\rangle + c_{A4}|11\rangle|h_{A4}\rangle)$$

$$U_1|1+\rangle|E_1\rangle = \frac{1}{\sqrt{2}}(c_{A1}|00\rangle|h_{A1}\rangle + c_{A2}|01\rangle|h_{A2}\rangle + c_{A3}|10\rangle|h_{A3}\rangle + c_{A4}|11\rangle|h_{A4}\rangle)$$

$$+ \frac{1}{\sqrt{2}}(d_{A1}|00\rangle|k_{A1}\rangle + d_{A2}|01\rangle|k_{A2}\rangle + d_{A3}|10\rangle|k_{A3}\rangle + d_{A4}|11\rangle|k_{A4}\rangle)$$

$$U_1|+1\rangle|E_1\rangle = \frac{1}{\sqrt{2}}(b_{A1}|00\rangle|g_{A1}\rangle + b_{A2}|01\rangle|g_{A2}\rangle + b_{A3}|10\rangle|g_{A3}\rangle + b_{A4}|11\rangle|g_{A4}\rangle)$$

$$+ \frac{1}{\sqrt{2}}(d_{A1}|00\rangle|k_{A1}\rangle + d_{A2}|01\rangle|k_{A2}\rangle + d_{A3}|10\rangle|k_{A3}\rangle + d_{A4}|11\rangle|k_{A4}\rangle),$$

Upon receiving the attacked qubits, Bob will measure the qubits in $S_A$ in The z-basis. And after he receives all qubits from Alice, he will check the hash value. If the attacker changes the qubits in The z-basis, the hash value check will fail, and then the attack will be detected. By the definition of robustness, we can assume the attacker will not change the value of the qubits in The z-basis. This gives a limit to the possible $U_1$, and we can denote the limited $U_1$ by:

$$U_1|0+\rangle|E_1\rangle = \frac{1}{\sqrt{2}}(|00\rangle|f_{A1}\rangle + |01\rangle|g_{A2}\rangle)$$

$$U_1|+0\rangle|E_1\rangle = \frac{1}{\sqrt{2}}(|00\rangle|f_{A1}\rangle + |10\rangle|h_{A3}\rangle)$$

$$U_1|1+\rangle|E_1\rangle = \frac{1}{\sqrt{2}}(|10\rangle|h_{A3}\rangle + |11\rangle|k_{A4}\rangle)$$

$$U_1|+1\rangle|E_1\rangle = \frac{1}{\sqrt{2}}(|01\rangle|g_{A2}\rangle + |11\rangle|k_{A4}\rangle).$$

When Bob sends the qubits back to Alice, Eve can use a different unitary operator $U_2$ to attack again. We denote the unitary operator $U_2$ by:

$$U_2|00\rangle|E_2\rangle = a_{B1}|00\rangle|f_{B1}\rangle + a_{B2}|01\rangle|f_{B2}\rangle + a_{B3}|10\rangle|f_{B3}\rangle + a_{B4}|11\rangle|f_{B4}\rangle$$

$$U_2|01\rangle|E_2\rangle = b_{B1}|00\rangle|g_{B1}\rangle + b_{B2}|01\rangle|g_{B2}\rangle + b_{B3}|10\rangle|g_{B3}\rangle + b_{B4}|11\rangle|g_{B4}\rangle$$

$$U_2|10\rangle|E_2\rangle = c_{B1}|00\rangle|h_{B1}\rangle + c_{B2}|01\rangle|h_{B2}\rangle + c_{B3}|10\rangle|h_{B3}\rangle + c_{B4}|11\rangle|h_{B4}\rangle$$

$$U_2|11\rangle|E_2\rangle = d_{B1}|00\rangle|k_{B1}\rangle + d_{B2}|01\rangle|k_{B2}\rangle + d_{B3}|10\rangle|k_{B3}\rangle + d_{B4}|11\rangle|k_{B4}\rangle,$$

where $|E_2\rangle$ is Eve's newly prepared quantum state and $|f_{Bi}\rangle, |g_{Bi}\rangle, |h_{Bi}\rangle, |k_{Bi}\rangle$ are Eve's quantum states after the attack.

The situations in the protocol are denoted by:

$$U_2 \frac{1}{\sqrt{2}}(a_{A1}|f_{A1}\rangle|00\rangle + b_{A2}|g_{A2}\rangle|01\rangle)|E_2\rangle$$

$$= \frac{1}{\sqrt{2}}(a_{A1}|f_{A1}\rangle)(a_{B1}|00\rangle|f_{B1}\rangle + a_{B2}|01\rangle|f_{B2}\rangle + a_{B3}|10\rangle|f_{B3}\rangle + a_{B4}|11\rangle|f_{B4}\rangle)$$

$$+ \frac{1}{\sqrt{2}}(b_{A2}|g_{A2}\rangle)(b_{B1}|00\rangle|g_{B1}\rangle + b_{B2}|01\rangle|g_{B2}\rangle + b_{B3}|10\rangle|g_{B3}\rangle + b_{B4}|11\rangle|g_{B4}\rangle)$$

$$U_2 \frac{1}{\sqrt{2}}(a_{A1}|f_{A1}\rangle|00\rangle + c_{A3}|h_{A3}\rangle|10\rangle)|E_2\rangle$$

$$= \frac{1}{\sqrt{2}}(a_{A1}|f_{A1}\rangle)(a_{B1}|00\rangle|f_{B1}\rangle + a_{B2}|01\rangle|f_{B2}\rangle + a_{B3}|10\rangle|f_{B3}\rangle + a_{B4}|11\rangle|f_{B4}\rangle)$$

$$+ \frac{1}{\sqrt{2}}(c_{A3}|h_{A3}\rangle)(c_{B1}|00\rangle|h_{B1}\rangle + c_{B2}|01\rangle|h_{B2}\rangle + c_{B3}|10\rangle|h_{B3}\rangle + c_{B4}|11\rangle|h_{B4}\rangle)$$

$$U_2 \frac{1}{\sqrt{2}}(c_{A3}|h_{A3}\rangle|10\rangle + d_{A4}|k_{A4}\rangle|11\rangle)|E_2\rangle$$

$$= \frac{1}{\sqrt{2}}(c_{A3}|h_{A3}\rangle)(c_{B1}|00\rangle|h_{B1}\rangle + c_{B2}|01\rangle|h_{B2}\rangle + c_{B3}|10\rangle|h_{B3}\rangle + c_{B4}|11\rangle|h_{B4}\rangle)$$

$$+ \frac{1}{\sqrt{2}}(d_{A4}|k_{A4}\rangle)(d_{B1}|00\rangle|k_{B1}\rangle + d_{B2}|01\rangle|k_{B2}\rangle + d_{B3}|10\rangle|k_{B3}\rangle + d_{B4}|11\rangle|k_{B4}\rangle)$$

$$U_2 \frac{1}{\sqrt{2}}(b_{A2}|g_{A2}\rangle|01\rangle + d_{A4}|k_{A4}\rangle|11\rangle)|E_2\rangle$$

$$= \frac{1}{\sqrt{2}}(b_{A2}|g_{A2}\rangle)(b_{B1}|00\rangle|g_{B1}\rangle + b_{B2}|01\rangle|g_{B2}\rangle + b_{B3}|10\rangle|g_{B3}\rangle + b_{B4}|11\rangle|g_{B4}\rangle)$$

$$+ \frac{1}{\sqrt{2}}(d_{A4}|k_{A4}\rangle)(d_{B1}|00\rangle|k_{B1}\rangle + d_{B2}|01\rangle|k_{B2}\rangle + d_{B3}|10\rangle|k_{B3}\rangle + d_{B4}|11\rangle|k_{B4}\rangle).$$

Alice will check all the received qubits. This check gives some limits of the possible of $U_1$ and $U_2$. Combine all limits above. We can get

$$U_1|00\rangle|E_1\rangle = |00\rangle|f_{A1}\rangle$$

$$U_1|01\rangle|E_1\rangle = |01\rangle|g_{A2}\rangle$$

$$U_1|10\rangle|E_1\rangle = |10\rangle|h_{A3}\rangle$$

$$U_1|11\rangle|E_1\rangle = |11\rangle|k_{A4}\rangle \ ,$$

$$U_2|00\rangle|E_2\rangle = |00\rangle|f_{B1}\rangle$$

$$U_2|01\rangle|E_2\rangle = |01\rangle|g_{B2}\rangle$$

$$U_2|10\rangle|E_2\rangle = |10\rangle|h_{B3}\rangle$$

$$U_2|11\rangle|E_2\rangle = |11\rangle|k_{B4}\rangle \ ,$$

And

$|f_{A1}\rangle = |g_{A2}\rangle = |h_{A3}\rangle = |k_{A4}\rangle$, $|f_{B1}\rangle = |g_{B2}\rangle = |h_{B3}\rangle = |k_{B4}\rangle$ .

We hence prove that there is no unitary operator to be used to get useful information about the pre-shared key and the newly shared key without being detected.

## 4 Comparison

**Table 1**. Comparison among measure-resend ASQKD protocols

|  | Yu et al.'s measure-resend ASQKD | Li et al.'s measure-resend ASQKD | The proposed measure-resend ASQKD |
|---|---|---|---|
| Classical participants' quantum capabilities | Measure Generate Reflect | Measure Generate Reflect | Measure Generate Reflect |
| Quantum resource | Bell state | Bell state, single photons | Single photons |
| Qubit efficiency | 1/10 | 1/9 | 1/6 |
| Required pre-shared keys (in bits) | $3n + 3m$ | $2n + 2m$ | $n + 2m$ |
| Required classical channel | Public discussion | 1-bit authenticated message | 1-bit authenticated message (2 times) |
| Hash function | No | Public hash | Secret hash |

To compare the proposed protocol with Yu et al.'s and Li et al.'s measure-resend protocols, (1) classical participants' quantum capabilities, (2) quantum resource, (3) qubit efficiency, (4) bit number of pre-shared keys, (5) classical channels required during key distribution, and (6) hash function are given in Table 1 above.

The qubit efficiency is defined as the number of distributed key bits divided by the total number of generated qubits [4]. It is assumed that $n = m$, then the qubit efficiency of the proposed protocol is $\frac{n}{(2n+2m) + (n+m)}$, which is equal to 1/6 and is higher than Yu et al.'s and Li et al.'s. $2n + 2m$ is for $Q_A$ sent from Alice, and $n + m$ is for photons belonging to $S_A$ resent from Bob.

# 5 Conclusions

Two-step quantum transmission is used in the proposed measure-resend ASQKD protocol to perform the key distributed from Alice to Bob. The process of the proposed protocol only needs two 1-bit authenticated classical messages. It is robust under the collective attack. The contributions of the proposed protocol are the following: (1) lowering down the burden of the quantum resource using single photons, (2) reducing the number of pre-shared keys, (3) providing a formal proof of security, and (4) achieving better qubit efficiency than the state-of-the-art measure-resend protocols.

## Acknowledgements

This research was partially supported by the Ministry of Science and Technology, Taiwan, R.O.C. (MOST 107-2627-E-006-001 and 108-2221-E-006 -107).